\documentclass[aps,prl,twocolumn,groupedaddress]{revtex4-1}
\usepackage{hyperref}

\newcommand{\f}{\frac}
\newcommand{\h}{\hspace{0.5mm}}

\newcommand{\ds}{\displaystyle}
\usepackage{color}

\begin{document}

\title{Implications Of A Dark Sector U(1) For Gamma Ray Bursts}

\author{Tom Banks}
\email{banks@scipp.ucsc.edu}
\affiliation{Department of Physics and SCIPP\\University of California at Santa Cruz, CA 95064.\\and\\Department of Physics and NHETC\\Rutgers University, Piscataway, NJ 08854}
\author{Willy Fischler}
\email{fischler@physics.utexas.edu}
\author{Dustin Lorshbough}
\email{lorsh@utexas.edu}
\author{Walter Tangarife}
\email{wtang@physics.utexas.edu}
\affiliation{Department of Physics and Texas Cosmology Center\\ The University of Texas at Austin,
TX 78712.}

\begin{abstract}
We discuss the implications for gamma ray burst studies of a dark unbroken $U(1)_D$ sector that couples predominantly through gravity to the visible sector.  The dominant dark matter component remains neutral under $U(1)_D$.  The collapsar model is assumed to explain the origin of long gamma ray bursts.  The main idea is that, by measuring the change in stellar black hole spin during the duration of the GRB, one can make inferences about the existence of a dark matter accretion disk.  This could potentially provide evidence for the existence for a $U(1)_D$ sector.
\end{abstract}

\maketitle
\section{Introduction}
The possible existence of a dark unbroken $U(1)_D$ sector, complete with dark photons, charged particles, and perhaps even dark Hydrogen has been extensively considered in the literature \cite{Holdom86,Demir:1999vv,Feng08,Ackerman:mha,ArkaniHamed:2008qn,Feng:2009mn,Kaplan09,Fischler10,Kaplan11,Cline12,Belostky:2012zz,Racine12,Randall12,Randall12b,Randall13,Fan13,Fischler14}.  For simplicity, we will consider dark matter that is neutral under the standard model.  The dominant dark matter component is neutral under the $U(1)_D$ sector.  We will assume there is some excess of dark charged particles such that neutral Hydrogen may be formed as in the visible sector.  In this type of scenario, dark matter may form more complicated astrophysical structure such as galactic disks \cite{Randall12,Randall12b,Randall13,Fischler14}.  Astrophysical observations such as halo shape analysis \cite{Buote02,Escude02,Peter12} and the bullet cluster \cite{Clowe03,Markevitch04,Bradac06,Randall08,McDermott10} will bound the amount of allowed charged dark matter, but ultimately cannot provide evidence for its existence.\\

We will assume that the collapsar model \cite{Woosley93,Paczynski98,MacFadyen98}, whereby the iron core of a progenitor star collapses into a black hole, describes some of the observed long gamma ray bursts and is followed by jet emission which is powered by the Blandford-Znajeck (BZ) mechanism \cite{BZ77,Komissarov01,Komissarov02,Komissarov2004,Koide2004,McKinney2004,McKinney2006,Barkov:2007us,Komissarov08,Barkov08,Nagataki2009,Komissarov09,Tchekhovskoy11, Penna13}.  We observe these jets as gamma ray bursts \cite{Barkov08}.  While there are competing theories about the origin of jets, numerical studies indicate that it is more likely that astrophysical jets are the result of the BZ mechanism rather than the Penrose mechanism or accretion disk braking \cite{Barkov:2007us,Barkov08,Komissarov09} , for at least some sets of parameters.\\

If a $U(1)_D$ sector exists, it will accrete around a black hole and emit jets of dark radiation that are unobservable by visible sector photodetectors.  The mechanism underlying both the dark jet production and the visible jet production is the same.  Therefore the black hole rotational energy will evolve differently than expected from the case of visible jets alone.  We derive an equation for the amount of $U(1)_D$ energy density in the vicinity of the newly formed black hole.  We expect the visible sector and dark sector accretion disks to be formed in a similar manner since they are subject to the same gravitational environment.  This will need to be checked with ``two-sector" numerical simulations.  We also note that the bounds we derive are easily evaded.
If the $U(1)_D$ charged dark matter is too sparse around the progenitor star, too massive or the dark fine structure constant too small, the effect disappears.

\section{Temporal Change in Black Hole Spin}
The extractable energy of a Kerr black hole is given by subtracting the irreducible mass contribution from the total energy \cite{MTW}
\begin{equation}\label{eq:E_rot}
E_{\text{extr}}=M_B\left[1-\f{1}{\sqrt{2}}\left(1+\sqrt{1-a^2}\right)^{1/2}\right],
\end{equation}
where $\ds{-1<a\equiv J/G\h M_B^2<+1}$ is the dimensionless spin parameter.  Defining $\ds{\gamma\equiv\left(1-\f{E_{\text{extr}}}{M_B}\right)}$ and solving (\ref{eq:E_rot}) for the spin parameter yields
\begin{equation}
a=\pm 2\h \gamma\h \left(1-\gamma^2\right)^{1/2}.
\end{equation}
Without loss of generality we will restrict ourselves to the case of prograde rotation in which $a$ is positive.  A similar analysis may be done for the case of retrograde rotation in which $a$ is negative.\\

The value of the spin parameter has recently been measured for several stellar mass black holes \cite{McClintock11}.  Our proposal is to constrain the $U(1)_D$ energy density in he vicinity of the progenitor star by measuring both the temporal change in stellar black hole spin and the energy emitted in visible jets during a gamma ray burst event.  Defining $\ds{\lambda\equiv -2\f{\left(1-2\h\gamma^2\right)}{\left(1-\gamma^2\right)^{1/2}}=2\sqrt{2}\left(\f{1-a^2}{1-\sqrt{1-a^2}}\right)^{1/2}}$, the derivative of the spin parameter with respect to time may be shown to be
\begin{equation}\label{eq:a_dot_0}
\dot{a}= 2\h \dot{\gamma}\h\f{\left(1-2\h\gamma^2\right)}{\left(1-\gamma^2\right)^{1/2}}\equiv-\lambda \h\dot{\gamma}.
\end{equation}

There are two contributions to $\dot{\gamma}$, one proportional to the change in rotational energy and the other proportional to the mass accretion rate
\begin{equation}\label{eq:gamma_dot}
\dot{\gamma}=-\f{\dot{E}_{\text{extr}}}{M_B}+\f{E_{\text{extr}}}{M_B}\f{\dot{M}_B}{M_B}\approx-\gamma\f{\dot{E}_{\text{extr}}}{M_B}.
\end{equation}
For the entirety of this study, we are interested in the time regime after an accretion disk has been formed so that accretion rate is a meaningful concept.  On the left hand side of (\ref{eq:gamma_dot}) we have assumed that the irreducible mass is constant (as is the case for a maximally efficient process \cite{BZ77}), $\ds{\dot{M}_{\text{ir}}=\dot{M}_B-\dot{E}_{\text{extr}}\approx0}$, and therefore all energy infall remains extractable over the duration of the burst event.  This approximation becomes better as the black hole spin parameter $|a|$ approaches unity.  Numerical studies have shown that the black hole spin rapidly grows during the collapsing stage \cite{MacFadyen98}.\\

Within the assumptions outlined in the introduction, the three contributions to the $\dot{E}_{\text{extr}}$ are infall of visible and dark matter, jet emission in the visible and dark sector (if the microscopic parameters in the dark sector allow for jet emission), and the emission of gravitational radiation
\begin{equation}\label{eq:Eextr}
\dot{E}_{\text{extr}}=\dot{M}_{\text{in,v}}+\dot{M}_{\text{in,D}}-L_{\text{jet,v}}-L_{\text{jet,D}}-L_{\text{gr}}.
\end{equation}
There may be other sources of energy loss that we are neglecting here which may be quantified with numerical simulations, but we will restrict our attention to these contributions.  Substituting (\ref{eq:gamma_dot}) and (\ref{eq:Eextr}) into (\ref{eq:a_dot_0}) we obtain our final expression for $\dot{a}$
\begin{equation}
\dot{a}=\f{\lambda\h\gamma}{M_B}\h\left(\dot{M}_{\text{in,v}}+\dot{M}_{\text{in,D}}-L_{\text{jet,v}}-L_{\text{jet,D}}-L_{\text{gr}}\right).
\end{equation}

If the measurement is consistent with $\dot{M}_{\text{in,D}}=L_{\text{jet,D}}=0$, there are several possible explanations.  Obviously, it may indicate that there is no such dark $U(1)_D$ sector in nature.  Or the mass-to-charge ratio is too large and/or the fine structure constant is too small so that pair production is not efficient \cite{BZ77} and/or the Alfven speed cannot exceed the local free fall speed at the ergosphere \cite{Komissarov09}.  In addition, the dark sector magnetic field does not benefit from the existence of the visible progenitor star magnetic field.  This could require a more efficient magnetic field generation by the dark accretion disk than is present in the visible accretion disk.\\

If the microscopic properties of the $U(1)_D$ sector are appropriate for observing an effect, it may indicate that there is not enough energy density of dark matter in the vicinity of the black hole to compete with the energy density of the visible sector.  The visible sector has former star remnants in the immediate vicinity to source visible jets.  The dark sector requires instead the presence of a dark structure which may be nearly coincident with the progenitor star.  In the scenario of \cite{Randall12,Randall12b}, existing constraints on MACHOS (Massive Compact Halo Objects) \cite{Moniez:2010zt,Iocco:2011jz} are not easily interpreted in a way that constrains the existence of these dark structures as is discussed in \cite{Randall12}.  We note that solar capture of dark matter in the visible progenitor star will not be significant since we have not allowed interactions between the dense core of the visible star and dark matter, other than gravitational interactions.  It has been shown that solar capture is inefficient in these types of models even if a small cross section with visible nucleons is allowed \cite{Fan13}.

\section{A Simple Estimate}
The order of magnitude estimates for the visible parameters are given by \cite{Barkov:2007us} $\ds{\f{L_{\text{jet,v}}}{M_B}\approx 10^{-4}/s}$, \cite{Bethe:1990mw,Komissarov09} $\ds{\f{\dot{M}_{\text{in,v}}}{M_B}\approx \left(\f{1\text{ s}}{1\text{ s}+\Delta t}\right)10^{-2}/s}$ and \cite{Scheidegger10} $\ds{\f{L_{\text{gr}}}{M_B}\lesssim 10^{-9}/s}$.  For any given observed system, simulations using the exact parameters of that system will need to be done to determine these values for that given system.\\

Observations have shown that long gamma ray bursts may easily last for $\sim10^2$ seconds.  We will be optimistic and assume spin measurements can be made over the timespan of $\sim10^2$ seconds starting roughly ten seconds after initial core collapse.  For simplicity we will assume that throughout the measurement the power emitted, $L_{\text{jet}}$, is constant.  In principle the power emitted will change with time, but this happens at a much slower rate than the spin parameter itself.  Therefore if the time between spin measurements do not allow the spin parameter to evolve significantly more than $1\%$, the variation of power emitted may be neglected for a rough estimate.\\

For an initial spin measurement (ten seconds after the initial collapse) of $a\approx0.8$, we obtain $\lambda\gamma\approx2.4$ and therefore $(\Delta a)\approx0.05$ at a time 100 seconds after collapse.  If dark matter infalls or emits jets this value may change.  To distinguish these two cases experimentally, we must have the precision to measure spin to at least the second decimal place with enough time resolution to distinguish the beginning and ending spin for a given visible jet event.

\section{Current and Future Measurements}
Recent measurements by NuSTAR, XMM-Newton, and Suzaku \cite{Brenneman13} of Fe K$\alpha$ spectral emission have allowed astronomers to fit the supermassive black hole spin $a$.  This method has also been employed for determining the spin of stellar mass black holes (see \cite{McClintock11} for a review and discussion of methods).  Since the redshift of supermassive black holes for which spin has been measured is comparable to that at which we observe some gamma ray bursts, it seems to us that it is possible in principle to perform this measurement for the stellar mass black hole that may underlie long gamma ray bursts.\\

The Astro-H \cite{Takahasi08,Takahasi10} experiment scheduled to launch in 2015 and proposed experiments such as IXO/AXSIO \cite{White10}, ATHENA+ \cite{Barcons12,Barret13}, Extreme Physics Explorer (EPE) \cite{Garcia11}, and the Large Observatory For Timing (LOFT) \cite{Feroci11} would further increase our abilities to measure these parameters to greater precision.  Further precision improvements on these values will greatly reduce the uncertainty in our calculation arising from ${\mathcal{O}}(1)$ factors.\\

In the previous section we showed that in order to detect the dark $U(1)_D$ sector one must be able to measure at least a $1\%$ change in the spin parameter with a time resolution capable of distinguishing the beginning and end of a long gamma ray burst.  To our knowledge, no attempt has been made to measure the spin of a stellar black hole during the gamma ray burst period.  We do not know whether current techniques and experiments used for measuring black hole spin are capable of doing so or not.  The purpose of this letter is to make the case that such measurements should be considered and would have profound implications.\\

Currently, the only evidence we have that dark matter exists is through gravitational effects, therefore it may be that dark matter only couples gravitationally to visible matter.  A discrepancy between how quickly a black hole decreases its spin and how much energy has been emitted in visible jets could provide positive evidence that there exists a dark unbroken $U(1)_D$ sector in the universe, which is capable of emitting jets of dark radiation.
\section*{Acknowledgments}
D.L. would like to thank Jimmy, Ioannis Keramidas, Pawan Kumar, Chris Lindner, John Craig Wheeler and Stan Woosley for helpful discussions.  We would like to thank Aditya Aravind, Alexander Tchekhovskoy and Serguei Komissarov for helpful comments on an earlier draft of this paper.  The work of T.B. was supported by the Department of Energy.  The work of W.F., D.L. and W.T. was supported by the National Science Foundation under Grant Number PHY-1316033.

\end{document}